\documentclass{article}%
\usepackage{amsmath}%
\usepackage{amsfonts}%
\usepackage{amssymb}%
\usepackage{graphicx}
\newtheorem{theorem}{Theorem}

\newtheorem{remark}[theorem]{Remark}

\begin{document}

\begin{center}
\textbf{A Note on} \textbf{StiffDNN} - \textbf{A DNN for Stiff Dynamic
Systems}

Wei Cai

Department of Mathematics

Southern Methodist University

Dallas, TX 75275

December 2, 2019

\bigskip

\textbf{Abstract}
\end{center}

In this note, we will present a specially designed deep neural network (DNN),
which will target components of the solution of different time rate
individually through perspective of the Laplace s-transform of the solution.
Each segment of s-frequency range will be approximated by a small size DNN
separately after a reference central rate is factored out, leaving the rest of
the frequencies confined within a short range of rates, and allowing fast and
uniform learning by normal DNN.


\section{Introduction}

Efficient and accurate solution of large dynamic systems arising from chemical
kinetics and neural networks have been an outstanding challenge in scientific
computing due to the large disparity in rate of changes, manifested in the
stiffness of the system. Many different numerical techniques have been
proposed over the decades to achieve fast and efficient time integration of
those systems, including implicit time discretization, stochastic simulation
algorithms, tau-leaping method, among other methods. In this note, we will
take advantage of the expressivity of deep neural network (DNN) to represent
the complicated dynamics of these systems. In order to handle the different
time scales intrinsic in these systems, we will propose a specially designed
DNN which will target components of the solution of different time rate
individually. In a previous work, a PhaseDNN was designed to handle wideband
learning by decomposing the data in the Fourier spectral domain into separate
frequency ranges, each of them will be phase-shifted to a low frequency range
centered at zero frequency first and be approximated by a small DNN. The
original data can be then learned once each of the low frequency version is
shifted back to its original frequency range.

Borrowing the same philosophy of the PhaseDNN, we will design the StiffDNN
through the Laplace transform domain, each segment of s-frequency range will
be approximated by a DNN separately after a reference rate is factored out,
leaving the rest of the frequencies confined within a short range of rates.

\section{StiffDNN for dynamics system}

Consider a typical nonlinear dynamics system with state variables

$\mathbf{y(t)}$ $\mathbf{=}$ $\mathbf{(}y_{1}(t),$ $y_{2}(t),$ $\cdots
,y_{n}(t))^{\intercal},n>>1,$%

\begin{equation}
\overset{\cdot}{\mathbf{y}}=\mathbf{f(}t,\mathbf{y),}\text{ \ \ \ }%
\mathbf{0\leq t\leq T.}\label{ode}%
\end{equation}

We will assume that for each $1\leq i\leq n,$ $y_{i}(t)$ can be expressed in
terms of M rates of decaying factor $e^{-\lambda_{i}t}$, i.e.,%

\begin{align}
y_{i}(t)  & =%
{\displaystyle\sum\limits_{k=1}^{K}}
[\chi_{\lbrack0,t_{k}]}(t)+H(t-t_{k})e^{-\lambda_{k}(t-t_{k})}]\phi
_{k}(t)\label{yi(t)}\\
& =%
{\displaystyle\sum\limits_{k=1}^{K}}
\chi_{\lbrack0,t_{k}]}(t)\phi_{k}(t)+%
{\displaystyle\sum\limits_{k=1}^{K}}
H(t-t_{k})e^{-\lambda_{k}(t-t_{k})}\phi_{k}(t)\nonumber\\
& \triangleq h_{i}(t)+%
{\displaystyle\sum\limits_{k=1}^{K}}
H(t-t_{k})e^{-\lambda_{k}(t-t_{k})}\phi_{k}(t),\nonumber
\end{align}
where the rates $\{\lambda_{k}>0\}_{k=1}^{K}$ are not ordered in magnitude
however, for a stiff system, we expect that $\lambda_{\max}/\lambda_{\min
}>>1,$   $\lambda_{\min}=\min_{k}\{\lambda_{k}\},\lambda_{\max}=\max
_{k}\{\lambda_{k}\}.$ The time instances $t_{k}$ are considered as the
\textit{unknown on-set time} of the decay of $\lambda_{k}$ rate and the number
of terms $K$ is unknown, depending the interaction among different components
of the reactant $y_{i}(t).\chi_{I}(t)$ is the characteristic function of the
interval $I$ and  $H(\cdot)$ is the Heaviside function. $\phi_{k}(t)$ will be a
slowly varying function in time.

Each term in (\ref{yi(t)}) will be approximated by some DNN:%

\begin{equation}
\phi_{k}(t)\sim V_{k}(t,\theta_{v}),\label{Vk}%
\end{equation}
where $V_{k}(t,\theta_{v})$ will be a small size DNN, and due to the
non-smoothness of $h_{i}(t),$ a PhaseDNN \cite{phaseDNN} will be used for its
approximation, namely%

\begin{equation}
h_{i}(t)\sim%
{\displaystyle\sum\limits_{l=1}^{L}}
e^{i\omega_{l}t}W_{l}(t,\theta_{w}),\label{Wk}%
\end{equation}
and similar, the discontinuous $H(t-t_{k})$ will be approximated by a PhaseDNN
\cite{phaseDNN},%

\begin{equation}
H(t-t_{k})\sim%
{\displaystyle\sum\limits_{l=1}^{L}}
e^{i\omega_{k,l}t}U_{k,l}(t,\theta_{u}).\label{Uk}%
\end{equation}
By plugging (\ref{Vk})-(\ref{Wk}) into (\ref{yi(t)}), we obtain the following
form of DNN approximation for $y_{i}(t)$%

\begin{equation}
\text{\ }y_{i}(t)=%
{\displaystyle\sum\limits_{l=1}^{L}}
e^{i\omega_{l}t}W_{l}(t,\theta_{w})+%
{\displaystyle\sum\limits_{k=1}^{K}}
{\displaystyle\sum\limits_{l=1}^{L}}
e^{is_{kl}t}U_{k,l}(t,\theta_{u})V_{k}(t,\theta_{v}),\label{stiff1}%
\end{equation}
where each of the DNNs is of small size and
\begin{equation}
s_{kl}=-\lambda_{k}+i\omega_{k,l}\label{skl}%
\end{equation}
can be viewed as the Laplace transform variable.

The form in (\ref{stiff1}) can be simplified into a compact form, defined as
the StiffDNN approximation,%

\begin{equation}
y_{i}(t)\sim%
{\displaystyle\sum\limits_{l=1}^{L}}
e^{i\omega_{l}t}W_{l}(t,\theta_{w})+%
{\displaystyle\sum\limits_{j=1}^{J}}
e^{is_{j}t}T_{j}(t,\theta_{t})\triangleq Y_{i}(t,\mathbf{\theta)}%
,\label{StiffDNN}%
\end{equation}
where $\mathbf{\theta=\{}\theta_{w},\theta_{t}\}$ consists of all the NN
parameters and%

\begin{equation}
\mathbf{y(t)}\sim\mathbf{Y(}t,\mathbf{\theta)=}\left(
\begin{array}
[c]{c}%
Y_{1}(t,\mathbf{\theta)}\\
\vdots\\
Y_{1}(t,\mathbf{\theta)}%
\end{array}
\right)  .
\end{equation}

\begin{itemize}
\item Loss function $l(\mathbf{\theta})$

We will use the residual of (\ref{ode}) for $\mathbf{Y(}t,\mathbf{\theta)}$
over some uniformly spaced time station $\{t_{s}\}_{s=1}^{S},$%
\begin{equation}
l(\mathbf{\theta})=%
{\displaystyle\sum\limits_{s=1}^{S}}
||\overset{\cdot}{\mathbf{Y}}\mathbf{(}t_{s},\mathbf{\theta)-f(}%
t_{s},\mathbf{Y(}t_{s},\mathbf{\theta))||}^{2}.\label{res}%
\end{equation}

\end{itemize}

\begin{remark}
(\textbf{Selection of} $\lambda_{k}$) As we do not know in aprior the rates in
(\ref{yi(t)}), so we will simply select a uniformly-space rates $\widetilde
{\lambda}_{k}$ within a prescribed range $[0,$ $\lambda_{\max}],$%
\[
\widetilde{\lambda}_{k}=k\Delta\lambda,\text{ \ }\Delta\lambda=\frac
{\lambda_{\max}}{K},\text{ \ }1\leq k\leq K.
\]
We will simply put $\widetilde{\lambda}_{k}$ in (\ref{skl}), namely,%
\begin{equation}
s_{kl}=-\widetilde{\lambda}_{k}+i\omega_{k,l}\label{skl_1}%
\end{equation}
At convergence of StiffDNN, each of the true $\lambda_{k}$ will be located
next to a $\widetilde{\lambda}_{k}$, such that
\begin{equation}
|\lambda_{k}-\widetilde{\lambda}_{k}|\leq\Delta\lambda,
\end{equation}
\ which implies that the rate-term  $\ $
\begin{equation}
e^{-\lambda_{k}(t-t_{k})}\phi_{k}(t)=e^{-\widetilde{\lambda}_{k}(t-t_{k}%
)}e^{-\left(  \lambda_{k}-\widetilde{\lambda}_{k}\right)  (t-t_{k})}\phi
_{k}(t)
\end{equation}
where $e^{-\left(  \lambda_{k}-\widetilde{\lambda}_{k}\right)  (t-t_{k})}$
will be a slowly varying function, which can be absorbed into $\phi_{k}(t)$
and approximated by the DNN in (\ref{Vk}).
\end{remark}

\section{Conclusion}

In this note, we have presented a new DNN for find the global in time solution
of stiff dynamics system. Numerical test of this idea will be given in a
future work.

\end{document}